\documentclass[aps,prd,twocolumn,nofootinbib,superscriptaddress]{revtex4}

\usepackage[utf8]{inputenc}	
\usepackage{amssymb}
\usepackage{amsmath}
\usepackage{graphicx}
\usepackage{subfigure}

\usepackage[dvips]{color} 
\begin{document}

\title{Thermal Mass limit of Neutron Cores}
\author{Zacharias Roupas}`
\affiliation{Institute of Nuclear and Particle Physics, N.C.S.R. Demokritos, GR-15310 Athens, Greece}
\email{roupas@inp.demokritos.gr}

\begin{abstract}
Static thermal equilibrium of a quantum self-gravitating ideal gas in general relativity is studied at any temperature, taking into account the Tolman-Ehrenfest effect. Thermal contribution to the gravitational stability of static neutron cores is quantified. The curve of maximum mass with respect to temperature is reported. At low temperatures the Oppenheimer-Volkoff calculation is recovered, while at high temperatures the recently reported classical gas calculation is recovered. An ultimate upper mass limit $M = 2.43M_\odot$ of all maximum values is found to occur at Tolman temperature $ T = 1.27mc^2$ with radius $R = 15.2km$. 
\end{abstract}

\maketitle

\section{Introduction}

Oppenheimer and Volkoff \cite{Oppenheimer:1939} calculated the upper mass limit of neutron cores, under self-gravity, assuming very cold, completely degenerate ideal fermion gas. Their result was $M_{OV} = 0.7M_\odot$. In a more realistic set-up that includes also some protons, electrons and muons in $\beta$-equilibrium, preventing neutron decay, was found that this upper mass limit \cite{Harrison:1958} is not affected. This value turned out to be very low compared to observations and was later, after the discovery of the first neutron star \cite{Hewish:1968}, regarded as a proof of the fact that nuclear forces have repulsive effects at supra-nuclear densities \cite{Cameron:1959,Zeldovich:1962}. When nuclear forces are taken into account the limit increases and may reach two and a half solar masses for cold cores depending on the model (see for example \cite{Haensel:2007,Lattimer:2012} and references therein). 	

In Ref. \cite{Roupas:2014sda}, I have calculated the maximum mass limit of ideal neutron cores considering the opposite case of Oppenheimer and Volkoff (OV) calculation, namely the non-degenerate case. This accounts for the relativistic classical ideal gas. Surprisingly, this classical limit $M_{cl} = 2.4M_\odot$ at radius $R_{cl} = 15.2km$, agrees perfectly with recent observations \cite{Antoniadis:2013pzd,Demorest:2010}. The corresponding temperature is so high that may only apply to hot protoneutron stars \cite{Burrows:1986me,Prakash:1997}.

Neutron stars are compact objects so dense that General Relativity becomes important. 
They are the remnants of core-collapse supernovae \cite{Baade:1934,Bethe:1985,PhysRevLett.58.1494,PhysRevLett.58.1490,RevModPhys.85.245}. They are composed \cite{Haensel:2007} of a dense, thick core, a thin crust and outer very thin envelope and atmosphere. The core determines the upper mass limit and the size of the star. It may be subdivided in the inner and outer cores. The inner core is ultra-dense with $\rho \geq 2\rho_N$ where $\rho_N = 2.8\cdot 10^{14} gr/cm^3$ is the normal nuclear density. Since these densities are unreachable from present laboratory experiments, its exact temperature, state of matter and therefore equation of state remains at present a mystery. Models vary \cite{Lattimer:2000nx,Haensel:2007} from (superfluid) $npe\mu$ gas, hyperons, Bose-Einstein condensates, kaons and pions to strange matter, deconfined quarks and quark-gluon-plasma. The outer core is consisted mainly of neutrons although some protons, electrons and muons are present that prevent neutron decay. The crust consists of heavy nuclei and near the matching region with the outer core, free neutrons are also present. The matching region is situated at density about \cite{Haensel:2007}
\begin{equation}\label{eq:rho_R_v}
	\rho_R \simeq \frac{\rho_N}{2} = 1.4\cdot 10^{14}\frac{gr}{cm^3},
\end{equation} 
where heavy nuclei can no longer exist. At this density a phase transition occurs towards the $npe\mu$ gas, through the capture of electrons by protons. 

As mentioned above, the core becomes unstable for masses higher than some limiting value, which depends on the equation of state. Most mass limits found in literature (see \cite{Haensel:2007} and references therein) are based on calculations at zero temperature, corresponding to the ultra-degenerate limit. These suggest that the core is stabilized against gravitational collapse mainly due to neutron-neutron nuclear forces and degeneracy pressure. However, at the birth stage of a neutron star, core-collapse is believed to be halted at extreme temperatures $kT/mc^2 \sim 0.05$ or more, where $m$ is neutron's rest mass, leading to the formation of the protoneutron star \cite{Burrows:1986me,Prakash:1997}. The protoneutron star acquires a neutron rich core and reaches the maximum temperature in seconds after the shock that follows the halt of collapse. Then, after a rapid cooling to ten or more times lower temperature within a minute, the neutron star's core continues to cool down for a long, depending on the model, period \cite{RevModPhys.64.1133,Page:2004fy,Yakovlev:2004iq}, to very low temperatures $kT/mc^2 \sim 10^{-7}$.

The maximum mass limit of the protoneutron star is crucial, because it is the one that actual determines if the star will enter the cooling phase or collapse into a black hole. Actually, it must be about $0.2M_\odot$ bigger than the observed cold neutron star's masses \cite{Prakash:1997}. 
Taking into account very recent observations \cite{Antoniadis:2013pzd,Demorest:2010}, that give an upper observed mass of $M_{obs} = 2M_\odot$ for neutron stars, it is evident that almost all(!) of the protoneutron star models in Prakash et al. \cite{Prakash:1997} are ruled out. However, more recently, old models were refined and new models appeared \cite{Benic:2014jia,Gruber:2014mna,Kojo:2014rca,Gandolfi:2014,Pearson:2014,Zappala:2014,PhysRevC.90.019904,PhysRevD.89.105027,PhysRevC.89.015806,doCarmo:2013,Fryer:2013,Thomas:2013,Katayama:2012} in order to account for the recently observed heavy neutron stars. Understanding the core structure of protoneutron stars is of great importance for Physics in general, because it will reveal the behaviour of matter at supranuclear densities and high temperatures that are unreachable from present laboratory experiments. Therefore, ruling out core models should not be taken lightly and extensive investigation is required. 

In this extensive work of Prakash et al. \cite{Prakash:1997}, maximum masses are calculated for a large number of models of the core's equation of state and for two temperatures for each model. It is evident in this work, that temperature only very slightly affects the maximum mass limit. However, the Tolman-Ehrenfest effect \cite{Tolman:1930,Tolman-Ehrenfest:1930} is not taken into account, so that heat's efficiency to support mass is underestimated. This effect accounts for the presence of a temperature gradient, because of thermal mass. If the effect is significant, which is reasonable at these high densities and temperatures, then many models currently ruled out could pass the observational test.

In present work, the Tolman-Ehrenfest effect is taken into account. In order to quantify the effect of heat, I consider the ideal gas approximation, neglecting completely nuclear forces as well as leptons that play a major role in protoneutron stars. This is an oversimplification, which, however, may provide useful insight on the roles of degenereracy and thermal pressure, as well as the neglected nuclear forces. In addition, at very high temperatures, the approximation of ideal gas seems reasonable. 
Maximum masses are calculated for the whole temperature regime from the OV to the classical calculation for a quantum ideal gas. So that, \textit{what is actually reported here, is the dependence on temperature of the Oppenheimer and Volkoff original calculation}, as in Figure \ref{fig:Mcr_logT}.

Another motivation of this work is purely theoretical, namely to understand the behaviour of Fermi-Dirac distribution, a pure quantum phenomenon, in relation to the Tolman-Ehrenfest effect, a pure relativistic phenomenon. Simply stated, this effect accounts for the fact that heat has mass and thus gravitates, leading to inhomogeneous temperature at thermal equilibrium. Is this in agreement with a quantum phenomenon such as the Fermi-Dirac distribution? These issues are addressed in sections \ref{sec:eos} and \ref{sec:te}. 

In section \ref{sec:ml} is calculated the critical maximum mass of ideal neutron cores with respect to temperature, including all relativistic effects. An ultimate upper limit, similar to the classical calculation
\begin{equation}\label{eq:thermal_limit}
	M_{max} = 2.43M_\odot
\end{equation}
corresponding to a Tolman temperature $k\tilde{T} = 1.27mc^2=1192MeV$ and radius
\begin{equation}
	R = 15.2km.
\end{equation} 
is reported for neutron cores matching with an outer crust. The existence of a specific temperature value, at which an upper mass limit occurs, seems significant and it was unexpected. Thermal pressure, including the Tolman-Ehrenfest effect, is dominant at high temperatures and increases drastically the maximum mass limit. It is also found, that masses of the order of the upper observational limit $M_{obs} = 2M_\odot$ can be sustained, even when nuclear forces are neglected, at temperature $k\tilde{T} = 0.19mc^2=178MeV$. Temperature values given here, is expected to be over-estimated. The reason is that if nuclear forces and leptons were taken into account, their stabilizing contribution would enable the core to sustain the same mass at lower temperature.

\section{The Equation of State}\label{sec:eos}

Let for the moment ignore General Relativity and work in the framework of Special Relativity. Recall the one particle energy distribution for a quantum ideal gas:
\begin{equation}\label{eq:quantum_dis}
	g(\epsilon) = \frac{1}{e^{\beta(\epsilon - \mu)}\pm 1} \;,\;
	\left\lbrace
	\begin{array}{l}
		(+)\;\mbox{for fermions} \\[2ex]
		(-)\;\mbox{for bosons}
	\end{array}
	\right.
\end{equation}
where $\epsilon$ is the energy of one particle, including rest mass in the relativistic case, and $\beta$, $\mu$ are the inverse temperature and chemical potential, respectively. Using the Juettner transforamtion 
\begin{equation}\label{eq:Juettner} 
	\frac{p}{mc} = \sinh\theta
\end{equation}
and the relativistic definition of energy \begin{equation}\label{eq:e_p}
	\epsilon = \sqrt{m^2c^4 + p^2 c^2},
\end{equation}
the distribution (\ref{eq:quantum_dis}) may be written in terms of $\theta$:
\begin{equation}\label{eq:1p_e}
	g(\theta) = \frac{1}{e^{-\alpha + b \cosh\theta }\pm 1} 
\end{equation}
where 
\begin{align}
\label{eq:b_def}
	&b \equiv \frac{mc^2}{kT} \\
\label{eq:alpha}
	&\alpha \equiv \frac{\mu}{kT}. 
\end{align}
In case of neutrons (and baryons or electrons), using the distribution (\ref{eq:1p_e}), the pressure $P$ and mass density $\rho$ may be written as \cite{Chandra:1938}:
\begin{align}
\label{eq:P_Q} 	
&P = \frac{8\pi m^4 c^5}{3h^3}\int_0^\infty \frac{\sinh^4\theta d\theta}{e^{-\alpha + b \cosh\theta }+1}
\\
\label{eq:rho_Q}
&\rho = \frac{8\pi m^4 c^3}{h^3}\int_0^\infty \frac{\sinh^2\theta \cosh^2\theta  d\theta}{e^{-\alpha + b \cosh\theta }+1}.
\end{align}
Thus, the equation of state is given in parametric form $P = P(\mu , T)$ and $\rho = \rho(\mu , T)$ by equations (\ref{eq:P_Q}) and (\ref{eq:rho_Q}).

The chemical potential is the amount of free energy 
\begin{equation}\label{eq:free}
F = E-TS
\end{equation} 
needed to give (or take) to (from) a system in order to add one particle under conditions of constant temperature.  If the system is receptive to adding particles (no external work needed) the chemical potential is negative, otherwise it is positive. In classical systems, adding one particle causes a huge entropy increase, since the number of available configurations increases greatly, contributing a big amount to the minus sign of equation (\ref{eq:free}) and therefore the chemical potential is negative. However, for quantum fermionic systems at low temperatures that are degenerate, adding one particle increases very slightly the entropy because the energy of the particle is certain. Due to the Pauli principle, it will occupy the highest available energy level, identified with Fermi energy in the completely degenerate case. The chemical potential is positive in this case. Thus, the parameter $\alpha$ shows the degree of degeneracy of the system. We have the following limits:
\begin{align}
\label{eq:limit_Q}		&\alpha \rightarrow +\infty : \mbox{ Ultra-degenerate limit} \\
\label{eq:limit_C}		&\alpha \rightarrow -\infty : \mbox{ Classical limit} 
\end{align}

The maximum possible mass of neutron gas that can be gravitationally bound without collapsing was calculated for the ultra-degenerate case (\ref{eq:limit_Q}) by Oppenheimer-Volkoff \cite{Oppenheimer:1939} who found $M_{OV} = 0.71M_\odot$ at $R = 9.5km$. I calculated the mass limit for neutron stars in the classical case (\ref{eq:limit_C}) in Ref. \cite{Roupas:2014sda} and found $M_{cl} = 2.43M_\odot$ at $R = 15.2km$. In this work is covered the whole range between these extreme cases for every $\alpha$ and hence every value of $T$ and $\mu$, using the equation of state in the parametric form (\ref{eq:P_Q}) and (\ref{eq:rho_Q}).

\section{Thermal Equilibrium in General Relativity}\label{sec:te}

Let restrict ourselves to the static spherically symmetric case in General Relativity for which the metric may be written in Schwartzschild coordinates as:
\begin{equation}
	ds^2 = g_{tt}dt^2 - g_{rr} dr^2 - r^2 d\Omega
\end{equation}
Since we consider an ideal gas, the energy-momentum tensor is the one of a perfect fluid:
\begin{equation}
	T^\mu_\nu = \mbox{diag}(\rho c^2,-p,-p,-p)
\end{equation}
In the presence of Gravity, the temperature and chemical potential, even at thermal equilibrium may depend on position due to the Tolman-Ehrenfest effect \cite{Tolman:1930,Tolman-Ehrenfest:1930}. Thus, in equations (\ref{eq:P_Q}) and (\ref{eq:rho_Q}), $T$ and $\mu$ are the proper temperature and proper chemical potential $T = T(r)$ and $\mu = \mu(r)$, respectively.

In Refs. \cite{Roupas:2014nea,Roupas:2013nt}, I have derived the conditions for thermal equilibrium in General Relativity for static spherically symmetric systems by extremizing the total entropy for fixed total energy and number of particles
\begin{equation}\label{eq:dS}
	\delta S - \tilde{\beta} \delta M c^2 + \alpha \delta N = 0.
\end{equation}
This condition leads to the Tolman-Oppenheimer-Volkoff (TOV) equation \cite{Tolman:1939,Oppenheimer:1939}:
\begin{equation}\label{eq:TOV}  
	\frac{d P}{dr} = - \left({\frac{P}{c^2}} + \rho\right) {\left(\frac{G\hat{M}}{r^2} + 4\pi G \frac{P}{c^2} r\right) \left(1 - \frac{2G\hat{M}}{rc^2} \right)^{-1} }
\end{equation}
assuming only the Hamiltonian constraint, which practically accounts for the mass equation 
\begin{equation}\label{eq:mass_d}  
 	\frac{d \hat{M}}{dr} = 4\pi \rho r^2,	
\end{equation}
where $\hat{M}(r)$ is the total mass (rest mass+thermal energy+gravitational field's energy) until point $r$ of the sphere. We denote $R$ the edge of the gas sphere and $M$ the total mass 
\begin{equation}
	M=\hat{M}(R). 
\end{equation} 
The Lagrange multiplier $\tilde{\beta}$ was proven in Ref. \cite{Roupas:2014nea} to be the Tolman inverse temperature $\tilde{\beta} = 1/k\tilde{T}$
\begin{equation}\label{eq:Tolman} 
	T(r)\sqrt{g_{tt}} \equiv \tilde{T} = const.
\end{equation}
The Lagrange multiplier $\alpha$ was found to be exactly the quantity of equation (\ref{eq:alpha}), namely:
\begin{equation}\label{eq:alpha_cond}
	\alpha \equiv \frac{\mu(r)}{kT(r)} = const.
\end{equation}
The differentiated form of equation (\ref{eq:Tolman}) may easily be calculated \cite{Roupas:2014nea}, using Einstein's equations, to be:
\begin{equation}\label{eq:tolman}  
	\frac{db}{dr} =  	-\frac{b}{P+\rho c^2} \frac{dP}{dr}, 	
\end{equation}
where $b$ is given by equation (\ref{eq:b_def}). 

Thus, at thermal equilibrium three conditions should hold: the TOV equation (\ref{eq:TOV}), the relation (\ref{eq:alpha_cond}) and Tolman relation (\ref{eq:tolman}).

Let me also note that thermal equilibrium implies a constant entropy per particle. The later determines the particle density distribution $n$ from the thermodynamic Euler relation
\begin{equation}\label{eq:euler}
	Ts = P + \rho c^2 - \mu n
\end{equation}
and conditions (\ref{eq:alpha_cond}) and (\ref{eq:tolman}). So that, equation $\frac{d}{dr}(s/n) =0$ gives using (\ref{eq:euler}), (\ref{eq:alpha_cond}) and (\ref{eq:tolman}):
\begin{equation}
	\frac{dn}{dr} = \frac{n c^2}{P + \rho c^2}\frac{d\rho}{dr}
\end{equation}

Tolman relation (\ref{eq:tolman}) expresses the fact that heat has mass and therefore gravitates. At equilibrium, a temperature gradient should form to balance the gravitational attraction of heat. Note that the `mass of heat' is found to play an analogous role also for cases out of equilibrium as pointed out in Ref. \cite{Herrera:2007tr}.

One question raised is if the Fermi-Dirac distribution (\ref{eq:P_Q}), (\ref{eq:rho_Q}) do satisfies the conditions of thermal equilibrium (\ref{eq:alpha_cond}), (\ref{eq:tolman}) and hence maximizes the entropy. Let us prove that the answer is affirmative.

We consider the definition (\ref{eq:alpha_cond}) for $\alpha$, namely $\alpha = \beta \mu$ and not $\alpha = \beta (\mu - mc^2)$, which would account for subtracting the rest mass. Note, that it is common to redefine the chemical potential for (special) relativistic systems by subtracting the particle rest mass energy from both particle energy (\ref{eq:e_p}) and the chemical potential, and thus leaving the quantum distribution function (\ref{eq:quantum_dis}) intact. However, in General Relativity this cannot be done, because the correct chemical potential has to satisfy equation (\ref{eq:alpha_cond}). 
Let 
\begin{equation}
	A = \frac{8\pi m^4 c^5}{h^3}e^{\alpha}
\end{equation} 
We assume 
\begin{equation}
	A = const.
\end{equation} 
and hence that the relation (\ref{eq:alpha_cond}) holds. 

We have using (\ref{eq:P_Q}) and (\ref{eq:rho_Q}), that:
\begin{equation}\label{eq:Prho}
	P + \rho c^2 = A \int_0^\infty \frac{\sinh^2\theta (\cosh^2\theta + \frac{1}{3}\sinh^2\theta) d\theta}{e^{-\alpha + b \cosh\theta }+1} 
\end{equation}
Let calculate $dP/dr$. We have:
\begin{equation}
 \frac{dP}{dr} = -\frac{1}{3}A\frac{db}{dr} \int_0^\infty \frac{e^{-\alpha+b\cosh\theta}\sinh^4\theta \cosh\theta d\theta}{\left(e^{-\alpha + b \cosh\theta }+1\right)^2}
\end{equation}
After one integration by parts it becomes:
\begin{align*}
 \frac{dP}{dr} &= \frac{A}{3}\frac{db}{dr} \int_0^\infty d\left(\frac{1}{e^{-\alpha + b \cosh\theta }+1}\right) \sinh^3\theta \cosh\theta  \\
	&= -\frac{1}{b}A\frac{db}{dr} \int_0^\infty \frac{\sinh^2\theta (\cosh^2\theta + \frac{1}{3}\sinh^2\theta) d\theta}{e^{-\alpha + b \cosh\theta }+1} 
\end{align*}
which by use of equation (\ref{eq:Prho}) gives
\[
	\frac{dP}{dr}= \frac{P+\rho c^2}{b}\frac{db}{dr}.
\]
Hence, Tolman relation (\ref{eq:tolman}) holds.  
The extension for Bose-Einstein distribution is trivial.

\section{Mass limits of Neutron Cores}\label{sec:ml}

We normalize mass density and pressure to the values:
\begin{equation}
	\rho_* = \frac{8\pi m^4 c^3}{h^3} \; ,\; P_* = \rho_* c^2.
\end{equation}
which suggests for length and mass the normalization:
\begin{equation}
	r_* = \sqrt{\frac{h^3}{32\pi^2 G m^4 c}} \; , \;
	M_* = r_* \frac{c^2}{G}.
\end{equation}
Particle mass $m$ determines only the scale and does not affect qualitatively the results. For neutrons, it is:
\begin{equation}
	r_* = 2.42km \; ,\; M_* = 1.64M_\odot \; ,\;\textstyle \rho_* = 1.83\cdot 10^{16}\frac{gr}{cm^3}
\end{equation}
Then $\rho$, $P$, $M$ and $r$ are measured in units of $\rho_*$, $P_*$, $M_*$ and $r_*$ so that, equations (\ref{eq:P_Q}), (\ref{eq:rho_Q}), (\ref{eq:mass_d}) and (\ref{eq:tolman}), using (\ref{eq:TOV}), are written as:
\begin{align}
\label{eq:P_ND} 	
&P = \frac{1}{3}\int_0^\infty \frac{\sinh^4\theta d\theta}{e^{-\alpha + b \cosh\theta }+1}
\\
\label{eq:rho_ND}
&\rho = \int_0^\infty \frac{\sinh^2\theta \cosh^2\theta  d\theta}{e^{-\alpha + b \cosh\theta }+1}
\\
\label{eq:TOV_ND}  
&\frac{d b}{dr} = - b {\left(\frac{\hat{M}}{r^2} + P r\right) \left(1 - \frac{2\hat{M}}{r} \right)^{-1} }
\\
\label{eq:mass_ND}  
&\frac{d \hat{M}}{dr} = \rho r^2,	
\end{align}
These equations (\ref{eq:P_ND}-\ref{eq:mass_ND}) form the system of equations we have to solve with initial conditions:
\begin{equation}
	b(0) = b_0 \; ,\; \hat{M}(0) = 0
\end{equation}
and the boundary condition (\ref{eq:rho_R_v}). This is achieved numerically by developing an appropriate computer program.

The system is solved for some value of $b_0$, by choosing these various pairs $(R,\alpha)$ that give the fixed edge density $\rho_R$ of equation (\ref{eq:rho_R_v}). This way an $M-R$ curve is drawn and the critical maximum mass corresponding to this $b_0$ is calculated. The process is repeated spanning the whole range of $b_0$ generating the values $M_{cr}(b_0)$. These correspond also to $M_{cr}(\tilde{T})$, since at each solution $(M,b_0)$ corresponds a Tolman temperature $\tilde{T}$. 

\begin{figure}
\begin{center}
	\includegraphics[scale = 0.55]{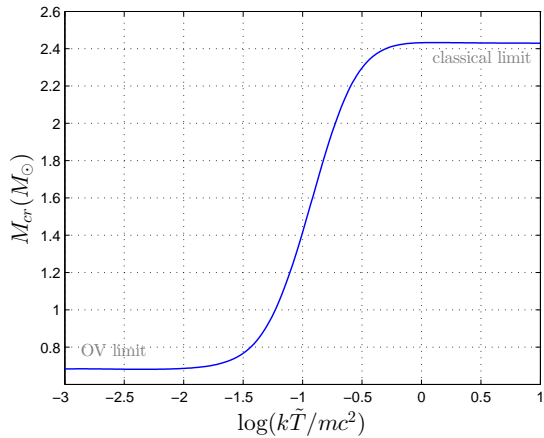}
	\caption{The critical maximum mass against gravitational collapse of ideal neutron cores with respect to the logarithm of Tolman temperature, where $m$ is the neutron mass. At low temperatures is recovered the OV calculation \cite{Oppenheimer:1939}, while at high temperatures the classical calculation \cite{Roupas:2014sda}. The mass limit increases rapidly in the temperature interval $k\tilde{T}/mc^2\in(0.01,1)$, where thermal energy gradually dominates over degeneracy pressure.
	\label{fig:Mcr_logT}}
\end{center} 
\end{figure}

\begin{figure}
\begin{center}
	\includegraphics[scale = 0.55]{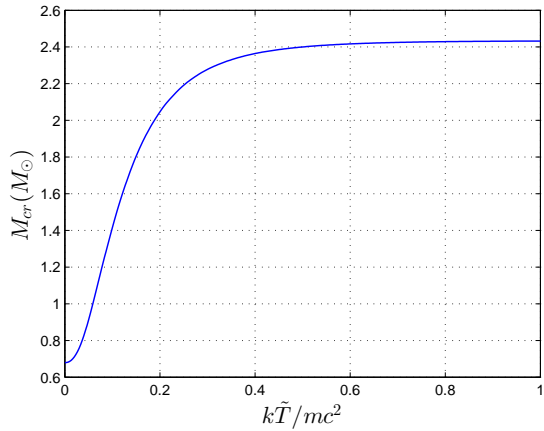}  
	\caption{The critical maximum mass with respect to Tolman temperature.
	\label{fig:Mcr_T}}
\end{center} 
\end{figure}

The critical mass with respect to $\log(k\tilde{T}/mc^2)$ is plotted in Figure \ref{fig:Mcr_logT}
 and with respect to $\tilde{T}$ in Figure \ref{fig:Mcr_T}. 
It demonstrates the dependence of OV limit on temperature. At low temperatures is recovered the OV calculation \cite{Oppenheimer:1939} and at high temperatures the classical calculation \cite{Roupas:2014sda}. It is evident that thermal energy enhances the ability of the system to sustain matter under self-gravity. It starts to dominate over degeneracy pressure at temperatures of the order $k\tilde{T} \sim 0.01 mc^2$. The mass limit is very rapidly increased from the OV limit to the classical limit in the temperature range $k\tilde{T}/mc^2 \in (0.01,1)$. 

Let me remark that, in fact, the critical mass in the case $\alpha\rightarrow +\infty$ corresponding to OV calculation is found here to be $0.68M_\odot$, a little lower than the value $0.71M_\odot$ reported in Ref. \cite{Oppenheimer:1939}. The reason is that, in the OV calculation the boundary condition is different. Namely they consider the vanishing of the pressure, while I consider the existence of a crust and a matching at the phase transition region between the core and the crust. So that, this $0.03M_\odot$ difference accounts approximately for the crust's mass. Assuming the vanishing pressure condition, the exact value is indeed recovered here. However, the pressure cannot be completely vanished at high temperatures. Most importantly, the matching with the crust is what actually happens. 

\begin{figure}
\begin{center}
	\includegraphics[scale = 0.55]{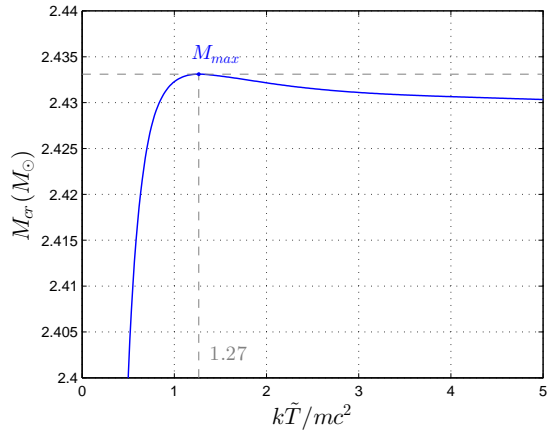}  
	\caption{The critical maximum masses of Figure \ref{fig:Mcr_logT} present a global maximum $M_{max} = 2.43M_\odot$ at $k\tilde{T} = 1.27mc^2$. At this maximum corresponds radius $R=15.2km$. It is also $\alpha \equiv \mu/k T = -8.89$, suggesting that the gas at these conditions is highly non-degenerate.
	\label{fig:M_max}}
\end{center} 
\end{figure}


A global maximum $M_{max} = 2.43M_\odot$ of the critical masses is found in agreement with Ref. \cite{Roupas:2014sda}. It occurs at $k\tilde{T} = 1.27mc^2=1192MeV$ and $\alpha = -8.89$, corresponding to a radius $R = 15.2km$ and a fractional redshift at the edge $z = 0.38$. This maximum is evident in Figure \ref{fig:M_max}. The $\alpha$ value suggests that the configuration at these conditions is highly non-degenerate. The temperature is too high for a protoneutron star, according to standard theory. However, as noted also in the Introduction, the inclusion of beta equilibrium and nuclear forces would normally enable the core to acquire the same mass limit at much lower temperature. 
Let me also note that the upper observed neutron stars mass \cite{Antoniadis:2013pzd,Demorest:2010}
$M_{obs} = 2M_\odot$
may be reached at temperature
$k\tilde{T} = 0.19mc^2 = 178MeV$
corresponding to the values $R = 14.4km$ and $\alpha = 0.31$.
This value of $\alpha$ suggests that the gas at these conditions is moderately degenerate, while the temperature value is much more realistic.

\begin{figure}
\begin{center}
	\includegraphics[scale = 0.55]{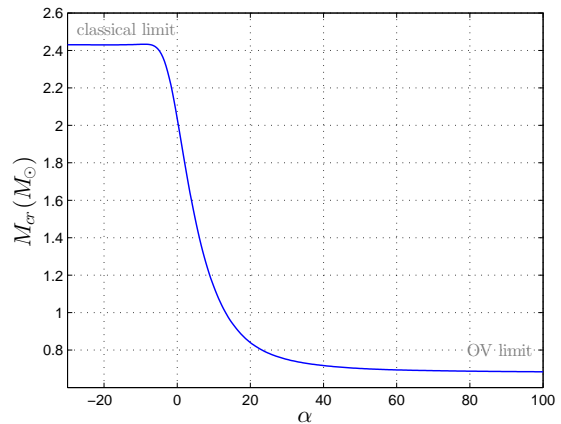}
	\caption{The critical maximum mass of ideal neutron cores with respect to the parameter $\alpha \equiv \mu(r)/kT(r)$ which is constant at thermal equilibrium. Parameter $\alpha$ determines the degree of degeneracy with $\alpha \rightarrow -\infty$ being the non-degenerate limit (classical calculation \cite{Roupas:2014sda}) and $\alpha \rightarrow +\infty$ the ultra-degenerate limit (OV calculation \cite{Oppenheimer:1939}).
	\label{fig:Mcr_a}}
\end{center} 
\end{figure}

In Figure \ref{fig:Mcr_a} is plotted the critical mass with respect to $\alpha = \tilde{\mu}/k\tilde{T}$, where is evident that $\alpha \rightarrow +\infty$ recovers the OV limit and $\alpha \rightarrow -\infty$ the classical calculation. 

\begin{figure}
\begin{center}
	\includegraphics[scale = 0.55]{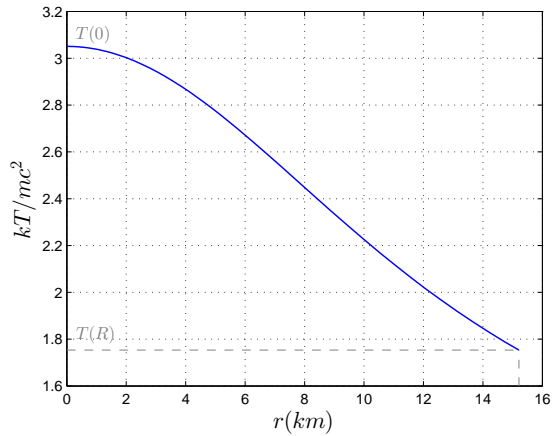} 
	\caption{The Tolman-Ehrenfest effect. Proper temperature with respect to radial coordinate for the marginal equilibrium corresponding to $M = M_{max} = 2.43M_\odot$ and Tolman temperature $k\tilde{T}/mc^2 = 1.27$.  Physically, Tolman temperature is the temperature at any point as measured by a distant observer. Thermodynamically, Tolman temperature is the variable conjugate to total mass-energy and therefore the temperature of the heat bath in the canonical ensemble.	
	\label{fig:Tolman}}
\end{center} 
\end{figure}
\begin{figure}
\begin{center}
	\includegraphics[scale = 0.55]{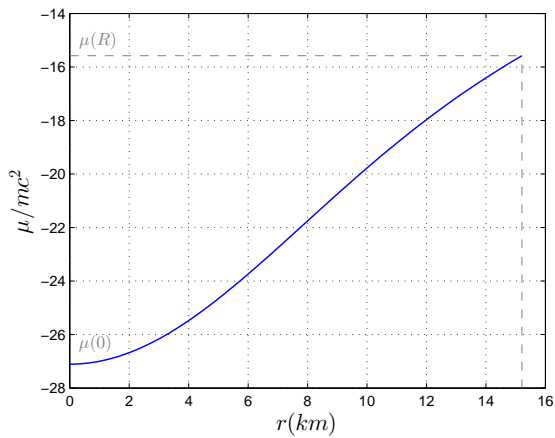} 
	\caption{The analogue of Tolman-Ehrenfest effect for the chemical potential. The proper chemical potential is plotted with respect to radial coordinate for the marginal equilibrium corresponding to $M = M_{max} = 2.43M_\odot$. It is $\tilde{\mu}/mc^2 = -11.3$ for this equilibrium. The quantity $\tilde{\mu}$ is the analogue of Tolman temperature and corresponds to chemical potential at any point as measured by a distant observer. 
	\label{fig:Tolman_mu}}
\end{center} 
\end{figure}

 In Figure \ref{fig:Tolman} is demonstrated the Tolman-Ehrenfest effect for the marginal equilibrium with $M = M_{max} = 2.43M_\odot$. The proper temperature $T(r)$ is the temperature measured by a local observer, i.e. it is the temperature realized by particles at $r$. Physically, the Tolman temperature $\tilde{T}$, which is a constant at thermal equilibrium, is the temperature measured at any point by a distant observer. Thermodynamically, it is the variable conjugate to total mass-energy and hence the temperature of the heat bath in the canonical ensemble. 

 In Figure \ref{fig:Tolman_mu} is demonstrated the Tolman-Ehrenfest effect for the chemical potential, at the marginal equilibrium with $M = M_{max} = 2.43M_\odot$. The proper chemical potential $\mu(r)$ is the chemical potential measured by a local observer, while $\tilde{\mu}$, which is constant at thermal equilibrium, is the analogue to Tolman temperature. 

\begin{figure}
\begin{center}
	\includegraphics[scale = 0.55]{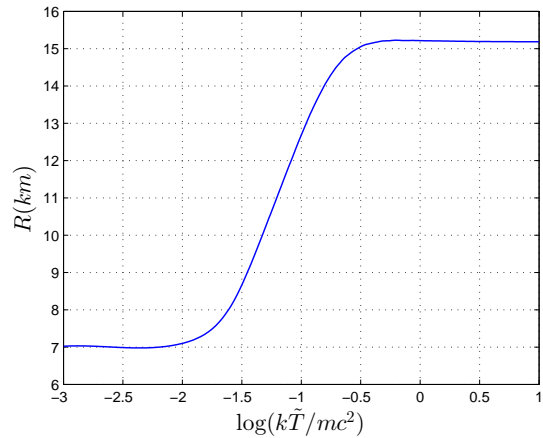} 
	\caption{The radius of neutron cores corresponding to each critical mass with respect to logarithm of Tolman temperature. It is evident that thermal energy increases the size of the core. Note that the ultra-degenerate limit gives lower result than the OV calculation $R_{OV} = 9.5km$. The reason is that in the OV calculation the core extends to the point of vanishing pressure, while here is considered to extend only until the crust. 
	\label{fig:R_logT}}
\end{center} 
\end{figure}

\begin{figure}
\begin{center}
	\includegraphics[scale = 0.55]{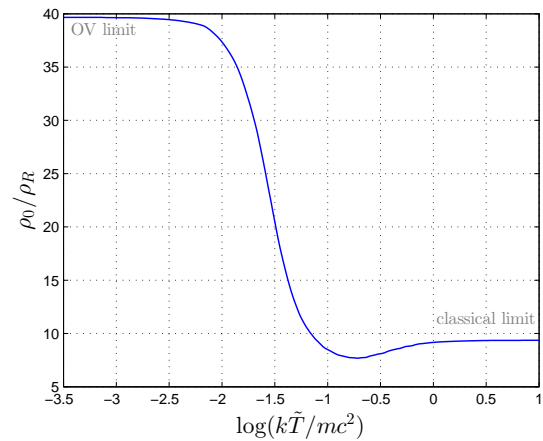} 
	\caption{The density contrast of neutron cores corresponding to each critical mass with respect to logarithm of Tolman temperature. It is evident that increasing the thermal energy tends to render the configuration more homogeneous. A global minimum equal to $\rho_0/\rho_R = 7.7$ appears at $k\tilde{T} = 0.19mc^2$, corresponding to critical mass $M_{cr} = 2.02M_\odot$.	
	\label{fig:DC_logT}}
\end{center} 
\end{figure}

In Figure \ref{fig:R_logT} is plotted the radius which corresponds to each $M_{cr}$ with respect to the Tolman temperature. Thermal energy inflates the gas, as one might expect, enabling it to acquire bigger radii. Also, mass is spread out tending to a more homogeneous state as is evident from Figure \ref{fig:DC_logT}. In this, is plotted the center to edge density ratio $\rho_0/\rho_R$ (density contrast) versus the Tolman temperature. There appears a global minimum $\rho_0/\rho_R = 7.7$ at $k\tilde{T} = 0.19mc^2$, corresponding to critical mass $M_{cr} = 2.02M_\odot$.	

The edge density cut-off $\rho_R$ does not affect drastically the critical mass at low temperatures, in which case most mass is concentrated in the center. But at high temperatures, the critical mass becomes more sensitive to the edge density cut-off, because matter is more homogeneously distributed. The dependence of the global mass maximum $M_{max}$ occurring at $k\tilde{T} = 1.27mc^2$ (see Figure \ref{fig:M_max}) with respect to the edge density cut-off $\rho_R$ and the particle mass $m_i$ is given by the following formula
\begin{equation}\label{eq:M_to_rho}
	M_{max} = 2.88\frac{m^2_n}{m^2_i}\sqrt{\frac{10^{14}\frac{gr}{cm^3}}{\rho_R}}M_\odot
\end{equation}
while the corresponding radius is
\begin{equation}\label{eq:R_to_rho}
	R|_{M_{max}} = 17.99\frac{m^2_n}{m^2_i}\sqrt{\frac{10^{14}\frac{gr}{cm^3}}{\rho_R}}km,
\end{equation}
where $m_n$ is neutron's rest mass.
I stress out, that the cut-off $\rho_R = \rho_N/2$, I used here, is not accidentally chosen, but it is the value at which a phase transition from the heavy nuclei gas to free neutrons gas occurs \cite{Haensel:2007}. I stress out that the core's mass limit of the present analysis does not depend on the equation of state for the crust, but only on the assumed minimum density for which a neutron gas can exist without forming nuclei, i.e. the cut-off $\rho_R$.

\section{Conclusions}

It is shown that quantum ideal self-gravitating gas is consistent with entropy extremization in General Relativity and in the static case. The quantum distributions are in accordance with the Tolman-Ehrenfest effect and all other conditions of thermal equilibrium, as is evident in sections \ref{sec:eos} and \ref{sec:te}. The corresponding formulation enabled us to perform a complete relativistic analysis of neutron ideal gas, taking into account all thermal effects.

The maximum mass limit, that an ideal neutron core can sustain without collapsing under self-gravity, is calculated with respect to temperature. In fact, it is reported here the actual dependence on temperature of the seminal Oppenheimer and Volkoff calculation \cite{Oppenheimer:1939}. At high temperatures is recovered the classical calculation of Ref. \cite{Roupas:2014sda}. 

It is evident that increasing thermal energy enables more massive cores to exist, rendering them also bigger and more homogeneous. These effects are quantified in Figures \ref{fig:Mcr_logT}, \ref{fig:R_logT} and \ref{fig:DC_logT}. An ultimate upper mass limit of all, temperature dependent, maximum masses appears at $k\tilde{T} = 1.27 m c^2=1192MeV$ with value $M_{max} = 2.43 M_\odot$, corresponding to radius $R = 15.2km$ and $\alpha = -8.89$. These mass and radius values agree perfectly with observations. The mass limit and corresponding radius depend on the edge density cut-off according to equations (\ref{eq:M_to_rho}), (\ref{eq:R_to_rho}). The corresponding temperature does not depend on edge density cut-off. 

The critical masses given here are expected to be realistic only at high temperatures and applying to hot protoneutron stars. However, since the final cold neutron star has less mass than the protoneutron star, due to cooling processes, the limits at high temperatures apply to cold neutron stars, as well. Therefore, the ultimate mass value reported here is an estimation of the maximum neutron stars mass.  

Comparing also the temperature and corresponding mass values of Figure (\ref{fig:Mcr_logT}) with the ones of Burrows \& Lattimer \cite{Burrows:1986me} and the ones of Prakash et al. \cite{Prakash:1997} --for pure neutron cores at high temperature (table 3 in this Ref. \cite{Prakash:1997})--  (see also \cite{Burgio:2006ed,Yazdizadeh:2011}) we see an approximate agreement, although in our case nuclear forces are completely neglected! 

Thus, thermal pressure, when Tolman-Ehrenfest effect is included, is equally or more efficient than nuclear forces, in halting gravitational collapse of neutron cores at high temperatures. Therefore, it is argued here, that it should not be neglected in calculation of protoneutron star maximum masses. 

This conclusion is also strengthened, by the finding that thermal, together with degeneracy pressure, even with nuclear forces neglected, can sustain the core at the upper mass $M_{obs}=2M_\odot$, suggested from actual observations, with radius $R = 14.4km$ at temperature $k\tilde{T} = 0.19mc^2=178MeV$. I note, that temperature values reported here are expected to be slightly over-estimated, because nuclear forces and leptons are neglected. If they were to be taken into account, they would contribute to outward pointing pressure, enabling the protoneutron star to stabilize at lower temperatures. 

Overall, this work suggests that neutron cores can be sustained against gravitational collapse at ultra high temperatures, because of heat, with masses comparable to the observable ones. The maximum neutron stars mass is estimated to be $2.4M_\odot$ with radius $15.2km$.


\bibliography{Roupas_Thermal_NS}
\bibliographystyle{h-physrev}

\end{document}